\documentstyle[12pt,blois,psfig]{article}

\begin{document}
\heading{THE IMPACT OF POPULATION III OBJECTS\\
         ON THE EARLY UNIVERSE}                    

\author{B. Ciardi $^{1}$, A. Ferrara $^{2}$} {$^{1}$ Dipartimento di
Astronomia, Florence, Italy.} {$^{2}$ Osservatorio Astrofisico di
Arcetri, Florence, Italy.}

\begin{bloisabstract}
We study the effects of the ionizing and dissociating photons produced
by Pop~III objects on the surrounding intergalactic medium. We find that
the typical size of a H$_2$ photodissociated region
is smaller than the mean distance between sources at $z \approx
20-30$, but larger than the ionized region. This implies that clearing
of intergalactic H$_2$ occurs before reionization of the universe is
complete. In the same redshift range, the soft-UV background in the
Lyman-Werner bands, when the intergalactic H and H$_{2}$ opacity is
included, is found to be $J_{LW} \approx 10^{-28}-10^{-26}$ erg
cm$^{-2}$ s$^{-1}$ Hz$^{-1}$. This value is well below the threshold
required for the negative feedback of Pop~III objects on the subsequent
galaxy formation to be effective. We have
combined these semi-analytical results with high resolution N-body   
simulations, to study the topological
structure of photoionization and photodissociation and the evolution of
the H$^+$ and H$_2$ filling factor.
\end{bloisabstract}

\section{Introduction}  
At $z\approx 1100$ the intergalactic medium (IGM) is expected to
recombine and remain neutral until the first sources of ionizing radiation
form and reionize it.  Until recently, QSOs were thought to be the main
source of ionizing photons, but observational constraints suggest the
existence of an early population of pregalactic objects (Pop~III
hereafter) which could have contributed to the reheating, reionization
and metal enrichment of the IGM at high redshift.

In order to virialize in the potential well of dark matter halos,
the gas must have a mass greater
than the Jeans mass, which, at  $z\approx 30$, is $\approx 10^{5}
M_{\odot}$, corresponding to virial temperatures 
$< 10^{4}$ K. To have a fragmentation of the gas
and ignite star formation, additional cooling is
required. It is well known that in these conditions the only efficient coolant
for a   plasma of primordial composition is molecular hydrogen
(\cite{AA}; \cite{T}). As the first stars form,
their photons in the energy range 11.26-13.6 eV are able to
penetrate the gas and photodissociate H$_{2}$ molecules both in the IGM
and in the nearest collapsing structures, if they can propagate that far
from their source, thus inhibiting subsequent formation of small mass
objects through the so called "negative feedback",
as Haiman, Rees \& Loeb (\cite{HRL}, HRL) have argued.
It is therefore important to assess the impact of these objects
on their surroundings through detailed calculations of the various
influence
spheres, {\it i.e.} ionization, photodissociation, and eventually also
supernova metal enriched (\cite{CF}) spheres, produced
by Pop~IIIs. In this talk we will briefly present the main results and
a progress report on these topics. A detailed derivation can be found in  
\cite{CFA}.

\section{Analytical estimates}        
If massive stars form in Pop~IIIs, their photons with $h\nu > 13.6$~eV
create a cosmological HII region in the surrounding IGM.
Its radius, $R_{i}$, can be estimated by solving the standard
equation for the evolution of the ionization front (\cite{DS};
\cite{SG}).
If steady-state is assumed and the cosmological
expansion is neglected (since $R_i \ll c/H$), then $R_i$ is approximately
equal to the Str\"omgren radius, $R_{S}$, that, in general, 
represents an upper limit
for $R_{i}$. For our reference parameters it is:
\begin{equation}
R_{i} \lsim R_{s} = 0.05 \, \left( {\Omega_b h^2} \right)^{-2/3}
(1+z)_{30}^{-2} \; S_{47}^{1/3}  \;\; {\rm kpc} ,
\label{rion}
\end{equation}
where $S_{47}=S_{i}(0)/(10^{47}$ s$^{-1}$) and $S_{i}(0)$ is the
ionizing photon rate.

In analogy with the cosmological HII region, photons
in the energy range 11.26 eV $\lsim h\nu \lsim$13.6 eV, create a
photodissociated sphere in the surrounding IGM. The main difference
with the ionization spheres
evolution is that there is no efficient mechanism to
re-form the destroyed H$_{2}$, analogous to H recombination. As a
consequence, it is impossible to define a photo-dissociation Str\"omgren
radius. However, given a point source that radiates $S_{LW}$ photons per
second in the LW bands, an estimate of the maximum
radius of the H$_2$ photodissociated sphere, $R_{d}$, is the distance at
which the photo--dissociation time becomes longer than the Hubble time:
\begin{eqnarray}\label{rcrit}
R_{d} \lsim 2.5 \, h^{-1/2} (1+z)_{30}^{-3/4} S_{LW,47}^{1/2} \;\; {\rm kpc},
\end{eqnarray}
where $S_{LW,47}=S_{LW}/(10^{47}$ s$^{-1}$). 

To substantiate the above analytical estimates we have developed a
non--equilibrium multifrequency radiative transfer code to
study the evolution of ionization and dissociation fronts 
produced by a point source of baryonic mass $M_{b}
\sim 10^{5} M_{\odot}$ forming at redshift $z=30$.
We have adopted a standard CDM model ($\Omega_m$=1, $h$=0.5,
$\sigma_8$=0.6), with $\Omega_b$=0.06.
The program evolves the energy equation and the chemical network
equations, including 27 chemical
processes and 9 species (H, H$^-$, H$^+$, He, He$^+$, He$^{++}$, H$_2$,
H$_2^+$ and free electrons). The cooling model includes, among other
processes, hydrogen, helium and molecular line cooling, Compton cooling on
the CMB, recombination cooling and all relevant photoionization
heating mechanisms.

\section{Results and implications}
\begin{figure}[t]
\centerline{
\psfig{figure=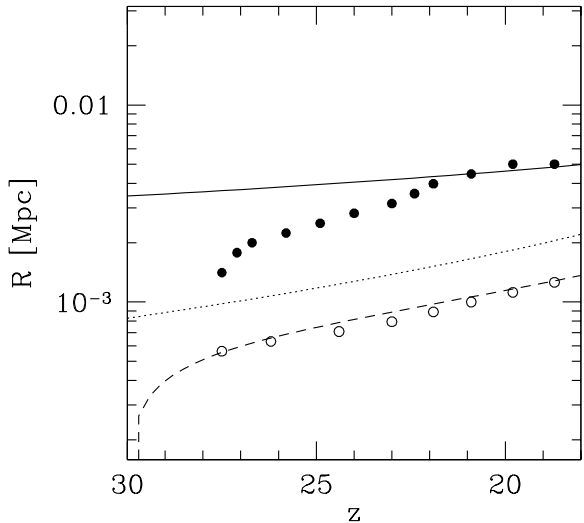,width=10cm}  
}
\caption{\label{fig1}
Ionization radius, $R_i$ (open circles) and
photodissociation radius, $R_d$ (filled)
of the regions produced by a Pop~III of total mass $M \approx 10^{6} \,
M_{\odot}$, turning on at $z \approx 30$, as a function of redshift.
Also shown is the maximum radius of the dissociated region (solid line),
given by eq.~(\ref{rcrit}), the Str\"omgren radius $R_{s}$ (dotted),
given by
eq.~(\ref{rion}), and the solution of the standard equation for the
evolution of the ionization front (dashed).
}
\end{figure}
In Fig.~\ref{fig1} we plot the numerical values of $R_{i}$ and
$R_{d}$ as function of redshift, for $\beta=1$, where $\beta$ is the
ratio between the flux produced by the object just below and above the
Lyman limit. Also shown are the upper limits to the radii, given by
eq.~(\ref{rion}) and eq.~(\ref{rcrit}) (with $S_{LW}=\beta S_{i}(0)$). 

To determine if the surviving intergalactic H$_2$ can build up a
non-negligible optical depth to LW photons, it is important to
compare the size of the dissociated regions around PopIIIs
with their average interdistance, $d\simeq [n(v_c,z)]^{-1/3}$, where
$n(v_c,z)$ is the proper number density distribution of dark matter
halos and $v_c$ is their circular velocity (\cite{PS}).
As in the redshift interval 20-30 $d$
is bigger than the typical derived $R_d$ ($d \approx 0.01-0.1$ Mpc),
the H$_{2}$ regions can hardly
overlap and completely destroy the primordial H$_2$ molecules, and the H$_2$
photodissociated sphere overlapping will become important at $z \le  20$.

\subsection{Soft-UV background}
In the calculation of the "soft-UV background" (SUVB), $J_{LW}$,
produced by PopIIIs, we have included the intergalactic
H$_{2}$ attenuation due
to the LW absorption lines, as well as the neutral H lines absorption.
The H lines are optically thick at
their center; this, combined with the effect of the cosmological
expansion, produces the typical sawtooth modulation of the spectrum.
We find that tipically, a SUVB intensity
$J_{LW} \approx 10^{-28}-10^{-26}$ erg cm$^{-2}$ s$^{-1}$ Hz$^{-1}$
is produced by Pop~III objects in the redshift range 20-30.

These results are particularly important when the effects of the
possible "negative feedback" are to be considered. HRL concluded that
in principle a SUVB created by pregalactic objects, would be able to
penetrate large clouds, and, by suppressing their H$_2$ abundance,
prevent the collapse of the gas. We find that the intensity of the SUVB is
well below the threshold required for the negative feedback to be effective.
Clearly, if at redshift $\approx 20$ complete overlapping of photodissociated
regions occurs, as previously suggested, the SUVB intensity can be
increased to
interesting values for negative feedback effects. However, by that time
a considerable fraction of the objects in the universe that must rely on
H$_2$ cooling for collapse might be already in
an advanced evolutionary stage and actively forming stars. To confirm
this hypotyhesis, that depends on the details of structure
formation, numerical simulations are required.

\section{Numerical simulation}
\begin{figure}[t]
\centerline{
\psfig{figure=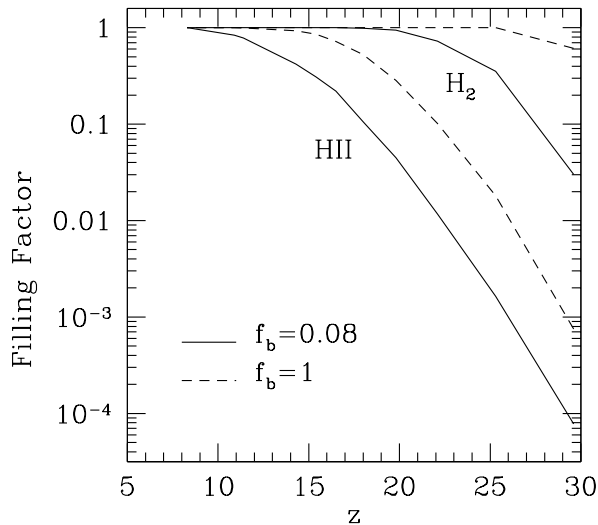,width=10cm}
}
\caption{\label{fig2}
Evolution of the H$_2$ (upper curves) and HII (lower curves) 
filling factor, for two different choices of $f_b$, the fraction of
virialized baryons able to cool and be available to form stars.
}
\end{figure}

We have been provided by F. Governato \& A. Jenkins of the Virgo
Consortium with very high-resolution N-body simulations (halos with masses
as low as $10^6 M_{\odot}$ are resolved) showing the dark matter halos
distribution from $z \approx 30$. The simulations have been performed by using
a P$^3$M parallel code, in a 5 Mpc box size (comoving) and
$253^3$ cells, for a CDM cosmology with
$\Omega_m=1,\, h=0.5$ and $\sigma_8=0.6$.
We have combined the numerical
results with the semi-analytical calculations, assigning to each halo the
ionized and dissociated spheres produced by the host PopIII, as 
computed above. This allows the study of the evolution of the
topological structure of photoionized/dissociated regions. As a preliminary
result, we show here the evolution of the filling factor (Fig.~\ref{fig2}).
As expected, clearing of H$_2$ occurs before reionization is completed.
On the other hand the reionization epoch ranges from $z\approx 8$ to 
$z\approx 12$ depending on the cooling efficiency $f_b$ of the collapsing
halos. 

\acknowledgements{We thank our collaborators in the project F. Governato,
A. Jenkins and T. Abel, for allowing presentation of unpublished work.}


%
\vfill

\begin{bloisbib}
\bibitem{AA} Abel, T., Anninos, P., Zhang, Y. \& Norman, M.L. 1997, {\it
NewA}, {\bf 2}, 181
\bibitem{CF} Ciardi, B. \& Ferrara, A. 1997, \apj {483} {L5}
\bibitem{CFA} Ciardi, B., Ferrara A. \& Abel, T. \apj {} {submitted}
\bibitem{DS} Donahue, M. \&  Shull, M.J. 1987, \apj {323} {L13}
\bibitem{HRL} Haiman, Z., Rees, M.J. \& Loeb, A. 1997, \apj {476} {458 (HRL)}
\bibitem{PS} Press, W. H. \& Schechter, P. 1974, \apj {187} {425}
\bibitem{SG} Shapiro, P.R. \& Giroux, M.L. 1987, \apj {321} {L107}
\bibitem{T} Tegmark, M. et al. 1997, \apj {474} {1}
\end{bloisbib}
\end{document}